\documentstyle[preprint,aps,prl,epsf]{revtex}

\begin{document}

\draft


\title{Partonic Energy Loss and the Drell-Yan Process}

\author{
G. T. Garvey$^a$ and J. C. Peng$^{a,b}$}
\address{
$^a$Los Alamos National Laboratory, Los Alamos, NM 87545\\
$^b$University of Illinois, Urbana, IL 61801
}
\date{\today}

\maketitle
\begin{abstract}
We examine the current status of the extraction of the rate of partonic energy loss
in nuclei from $A$ dependent data. The advantages and difficulties of using the 
Drell-Yan process to measure the energy loss of a parton traversing a cold nuclear 
medium are discussed. The prospects of using relatively low energy proton beams 
for a definitive measurement of partonic energy loss are presented.

\end{abstract} 

\eject

Characterizing the energy loss of partons in nuclear matter has been a
matter of considerable 
theoretical~\cite{gyulassy94,zakharov96,baier97,baier98,baier01,yang98,guo00,johnson02} and 
phenomenological~\cite{baier97,baier98,baier01,johnson02,wang98,vasiliev99,johnson01,arleo02,wang02}
interest over the past few years. In spite of this activity there is no
agreed to value for partonic $dE/dx$ (GeV/fm) in either 
cold or hot nuclear matter,
though there is general agreement that the rate of energy loss in the
latter exceeds that of the former.   

Much of the problem stems from the inability to directly measure this
energy loss, and the lack of a common agreement of the
processes and mechanisms to be included in specifying the energy loss. 
The purpose of this
communication is to briefly present the current status of the problem,
present a clear idea of quark energy loss in the nuclear medium, and
delineate an experimental program capable of determining the parameters
to adequately characterize quark energy loss in cold nuclei. We do not deal
with the issue of energy loss in relativistic heavy ion collisions (hot
nuclear matter), because the complexity of these reactions makes
quantitative extraction of partonic energy loss highly uncertain with
present knowledge. We will further adhere to the conventional assumption
that the rate of energy loss is independent of parton momentum at 
sufficiently high 
energies~\cite{gyulassy94,zakharov96,baier97,baier98,baier01}, 
though this assumption has been questioned~\cite{brodsky93}.

Table I demonstrates the lack of agreement among the extracted values for
quark energy loss that exist at the present time. Admittedly, much of
the difficulty may be matters of definition. There is variability in the
implicit definitions of what is included in the energy loss and the
correct path length over which the energy loss takes place. To reduce
some of the confusion in Table I, an average path length is employed
using the conventional value, $\langle L \rangle = 3/4~(1.2~A^{1/3})$ fm.
There are more
elaborate treatments~\cite{johnson02,johnson01} of the nuclear 
path length that yield appreciably shorter path lengths and hence 
higher rates of energy loss.

Doubtless some of the discrepancies in Table I would be reduced if
consistent definitions, analysis and proper attention to systematic
errors were applied. The nuclear dependence of fragmentation appears to
us to be too uncertain at the present time to permit a reliable
extraction of partonic energy loss. The $A$ dependence of the $p_T$ broadening
of the $p-A$ Drell-Yan (D-Y) yield may be an excellent measure of the
incident quark energy loss due to gluon radiation but the small value
observed for the broadening indicates that non-perturbative physics
is involved~\cite{alde90,pat99}. The above procedures will not
take account of the significant 
energy loss ($\sim 1$ GeV/fm) 
due to the color confining interaction which is operative even in vacuum.
The same criticism may be levied against any direct
comparison of jet energy loss in lepton-nucleus to lepton-proton DIS.

We believe that the best measure of the quark energy loss in cold
nuclear matter is the nuclear dependence of the $p-A$ D-Y
cross section. Viewed in the infinite momentum frame the energy loss of
the incident quark is terminated within the nuclear medium by the $q - \bar q$
annihilation process. This approach has been taken by 
refs.~\cite{johnson02,vasiliev99,johnson01,arleo02}
shown in Table I. The extracted energy losses are seen to be discrepant and
are discussed below. First, for illustrative purposes a simple description 
of how nuclear energy
loss affects the measured D-Y yield is presented. The $p-A$, leading-order
D-Y cross section per nucleon is given by  

\begin{eqnarray}
\frac {d\sigma^{DY}_{p-A}}{dx_1dx_2} = \frac{4\pi \alpha ^2_{em}}
{9 s} \frac {x_1 x_2} {x_1 +x_2} \sum_{i} e^2_i [q^p_i(x_1) \bar
q^A_i(x_2) + 
\bar q^p_i (x_1) q^A_i(x_2)].
\label{eq:eq1}
\end{eqnarray}

\noindent The fractional momenta $x_1$ and $x_2$ of the 
partons of flavor $i$ in the beam and target
respectively are determined from the mass of the lepton pair, 
$M^2 = s x_1 x_2$,
combined with the c.m. longitudinal momentum ($p_l$) of the pair
$x_1 - x_2 = 2 p_l/\sqrt{s}$.
For large $x_1$ ($x_1>0.5$) and for an interval in $x_2$ where 
$\bar q^A(x_2)/ \bar q^D (x_2) \sim 1$, which occurs for $0.1
< x_2 < 0.3$ as shown in~\cite{alde90}, the ratio of the D-Y
yields from nucleus $A$ to that from deuterium can be expressed as

\begin{eqnarray}
(\frac {d\sigma^{DY}_{p-A}}{dx_1dx_2})/(\frac {d\sigma^{DY}_{p-D}}{dx_1dx_2})
 \sim \frac{q^p_u(x^A_1)}{q^p_u(x^D_1)}
\sim \frac{(1-x^A_1)^3}{(1-x^D_1)^3}.
\label{eq:eq2}
\end{eqnarray}

\noindent The cubic dependence on $(1-x_1)$ comes from quark counting 
rules which are approximately correct as $x_1 \to 1$.
The energy of the incident quark is equal to $E_p x_1$.
Assuming that the incident quark experiences an energy loss per unit
length, $\alpha$, in
nuclear matter, and that there is no such energy loss in deuterium, then

\begin{eqnarray}
x^D_1 = x^p_1,~~~x^A_1 = x^p_1 + \alpha \frac{\langle L \rangle_A}
{E_p},
\label{eq:eq3}               
\end{eqnarray}

\noindent where $\langle L \rangle_A$ is the average path length of 
the incident quark in the
nucleus $A$. Because of the energy loss, $x^A_1$ must originate from larger
values of $x^p_1$ than is the case for deuterium. Eq.~\ref{eq:eq2} 
can then be expressed to lowest order in $\langle L \rangle_A / E_p$ as

\begin{eqnarray}
(\frac {d\sigma^{DY}_{p-A}}{dx_1dx_2})/(\frac {d\sigma^{DY}_{p-D}}{dx_1dx_2})
 \sim \frac{(1-x^p_1-\alpha\frac{\langle L \rangle_A}{E_p})^3}
{(1-x^p_1)^3} \sim 1 - \frac {3 \alpha \langle L \rangle_A}{E_p (1-x^p_1)}.
\label{eq:eq4}
\end{eqnarray}

\noindent Eq. 4 shows that the effects of partonic energy
loss are the largest at lower proton energies and at larger
$x_1$. Under readily achievable conditions with $A=200$, 
$E_p = 120$ GeV
and $x_1 = 0.7$, Eq.~\ref{eq:eq4} yields 0.78 
for $\alpha$= 0.5 GeV/fm, a readily measurable 
suppression due to
partonic energy loss. Of course, for a more quantitative analysis the 
approximations used in this paragraph should not be employed. 
The features presented in Eq.~\ref{eq:eq4} 
remain when measured parton distributions are employed.

Why then, is there such difficulty in extracting a reliable value for
nuclear partonic energy loss? First, there has not been experimental activity 
focused on elaborating partonic energy loss. The
available data come from a variety of sources~\cite{vasiliev99,alde90,badier81}
that focused
on other aspects of nuclear dependence in the D-Y process.  There is the
additional serious problem of separating the effects of shadowing from those of
energy loss. 
We recall that nuclear shadowing effect was observed in Deep Inelastic
Scattering for $x$ less than $\sim$0.07. This effect is interpreted in
the infinite momentum frame as a reduction of parton density at
small $x$ due to recombination of quarks and gluons from
different nucleons. In the target frame, nuclear shadowing occurs
when the coherence length, $L_c\approx 1/(2m_Nx_2)$, grows larger than the
average distance between nucleons. Both the energy loss and the shadowing
effects lead to an $A$-dependent reduction of the D-Y
cross section. In principle, they are readily separable as shadowing
reduces the yield at small values of $x_2$ while the effects of energy loss
are expected to be most pronounced for large values 
of $x_1$. However, the acceptance of
typical fixed-target detectors are biased to favor small values of $x_2$
in conjunction with large values of $x_1$. 
References~\cite{johnson02,vasiliev99,johnson01}
in Table I represent 
different analyses of the nuclear dependence of $E_p = 800$ GeV,
$p-A$ D-Y yield. Reference~\cite{vasiliev99} used $A$-dependent data 
from FNAL E866
which had had a large fraction of its data at $x_2 < 0.05$ and thus
required careful attention to nuclear shadowing. The shadowing corrections 
employed a phenomenological shadowing scheme due to Eskola 
et al.~\cite{eskola98}
which had determined the shadowing of sea quarks using the FNAL E772 data
with no account of the effects of parton energy loss.
Such an analysis naturally produces the 
small energy loss shown in Table I.

References~\cite{johnson02,johnson01} analyzed 
data from FNAL E772~\cite{alde90} and
FNAL E866. They utilized a very different reference frame and 
prescription for
calculating shadowing than used in ~\cite{vasiliev99}. 
Specifically, the nuclear
rest frame was employed so the lepton pair originates from a heavy
photon bremsstrahlunged off the incident fast charged quark. This description
has appreciably less shadowing and consequently requires greater parton
energy loss to account for the reduced D-Y yield with increasing $A$.  
Reference~\cite{johnson02} has a 
complete description of this analysis procedure 
and demonstrates that it correctly accounts for shadowing in DIS.
The energy loss per unit length reported in ~\cite{johnson02,johnson01} 
is more than twice as large as that shown in Table I 
because in ~\cite{johnson02,johnson01} the
parton's path length, $\langle L \rangle_{A=184}$ in Tungsten is only 2.1 fm.
This rather short parton path length is due to the requirement 
in~\cite{johnson02,johnson01} that the incident proton needs to travel
a certain distance (mean-free-path $\sim$ 2.5 fm) before an interaction
can liberate the high-$x$ parton.
The results shown in Table I use 
the commonly defined parton path length
to make the results more comparable. In~\cite{johnson02,vasiliev99,johnson01},
the incident proton energy is high 
($E_p$ = 800 GeV) so that
fractional energy loss is small, making the extracted energy loss
strongly dependent on the applied shadowing corrections. 

The strongest evidence for a small rate of parton energy loss would appear to 
come from the nuclear dependence of $\pi^- - N$ D-Y production.
Two experiments, one at 225 GeV/c~\cite{anderson79} and one at
150 GeV/c~\cite{badier81} reveal little in the way of nuclear dependence.
An analysis~\cite{arleo02} of the data from~\cite{badier81} 
is shown in Table I.
The value obtained, $-dE/dx = 0.20\pm 0.15$ GeV/fm, is in serious 
disagreement with [8,11]. Apart from explanations involving the low
statistics of~\cite{badier81} and incomplete knowledge of the pion's
parton distribution, the difference might lie with a smaller $\langle L
\rangle_A$ for partons from the pion. The $\pi$-N cross section is only
$60\%$ that of N-N resulting in a longer nuclear mean-free-path for
the pion and a corresponding shorter $\langle L \rangle_A$ for the partons.

Thus, due to the multiplicity of significant physical effects and the sparse 
database available, there is not yet a reliable value for the
partonic energy loss in cold nuclei. To remedy this situation we propose
a focused set of $p-A$, D-Y measurements at $E_p \sim 100$ GeV. 
The lower energy
enhances the effects of energy loss, accesses a region of $x_2$ where
shadowing is negligible, and using a proton beam allows the accumulation
of adequate statistics in a reasonable time. The use of several targets
of differing $A$ will be necessary in order to extract the effects of the
appropriate parton path length, as well as the relative contributions of
linear and quadratic dependence  of the energy loss on path length. 

To assess this issue more quantitatively we have examined the projected
sensitivity for D-Y experiments using parton distributions fit to the
world data and have incorporated nuclear effects in addition to
parton energy loss. In particular, we consider using the 
proton beams from both 
the 120 GeV Fermilab Main Ring Injector and the future 50 GeV Japan Hadron
Facility. Fig. 1 shows how parton energy loss
would affect the $(p+W)/(p+d)$ per nucleon D-Y yields at these beam energies.
In this calculation, we use the kinematic acceptance of a dimuon spectrometer
proposed for Fermilab Experiment 906~\cite{E906} and a similar spectrometer
considered~\cite{JHF} at the Japan Hadron Facility.
Only D-Y muon pairs with $M > 4.2$ GeV are used to avoid contamination from 
charmonium decays. The coverages in $x_2$ are $0.1 < x_2 < 0.35$ and
$0.2 < x_2 < 0.5$, respectively, for beam energies of 120 GeV and
50 GeV, well outside the small-$x_2$ region of nuclear shadowing.
The parton distribution
functions are taken from~\cite{mrst} and their modifications in the nuclear
medium are from~\cite{eskola98}. Leading-order D-Y formula given in Eq. 1
was used in this calculation. The effect of the next-to-leading order
diagrams is to introduce an overall normalization factor which
is expected to cancel in the D-Y cross section ratios.
The average path length in nucleus was taken as
$3/4~(1.2~A^{1/3})$. The various curves in Fig. 1
correspond to a partonic energy loss, $-dE/dx$, of
0.0, 0.1, 0.25 and 0.5 GeV/fm, respectively.
As expected, these curves clearly demonstrate the great sensitivity of
the D-Y cross section ratios to partonic energy loss when the incident 
proton energy is as low as shown in these examples. The projected statistical
errors for a 60-day run on each target are also shown in Fig. 1.
A partonic energy loss as small as 0.1 GeV/fm is seen to be readily measurable.

$A$-dependent D-Y data can further determine whether the
energy loss is linear or quadratic with the path length.
This is illustrated in
Fig. 2 where D-Y cross section ratios are shown as a function
of $A$. The solid circles correspond to a linear 
energy loss  of 0.25 GeV/fm.
The open squares correspond to the case where the energy loss is
proportional to the path length squared and matched to give
the same energy loss when $A=184$ (Tungsten target).
Fig. 2 shows that one can easily distinguish between $L$ and
$L^2$ dependence of the energy loss even when it is as small as 0.25
GeV/fm.

Thus it appears that many of the presently confusing issues of
partonic energy loss can be resolved via focused measurement of the p-A
dependence of the Drell-Yan cross section. A comparison of the yields
from several nuclear targets relative to that from deuterium should 
determine the amount of energy loss and the appropriate path length,
thereby firmly establishing the rate of energy loss. In addition to 
solving an interesting problem in its own right, these measurements will
have a direct impact on nuclear shadowing and on
the interpretation of many hard scattering processes in nuclei.

\eject

\begin{table}[tbp]
\caption {List of some published rates of parton energy loss in nuclear matter.}
\begin{center}
\begin{tabular}{|l|l|l|}
Reference & $-dE/dx$ (GeV/fm) & Data and Analysis Employed \\
\hline
\cite{wang98} & $< 0.02$ & $Pb-Pb$ jet quenching at SPS \\
\hline
\cite{baier98} & $\approx 2.8$ & $p_T$ broadening of $p-A$ jets \\
\hline
\cite{baier97} & $\approx 0.4$ & $p_T$ broadening of $p-A$ D-Y yield\\
\hline
\cite{guo00} & $\approx 1.2$ & Nuclear modification of $e-A$ fragmentation functions \\
\hline
\cite{wang02} & $\approx 0.5$ & Nuclear modification of $e-A$ fragmentation functions \\
\hline
\cite{vasiliev99} & $< 0.44$ & Nuclear dependence of 800 GeV $p-A$ D-Y cross sections \\
\hline
\cite{johnson02} & $1.12 \pm 0.15 \pm 0.21$ & Nuclear dependence of 800 GeV  $p-A$ D-Y cross sections \\
\cite{johnson01} & $0.95 \pm 0.21 \pm 0.21$ & \\
\hline
\cite{arleo02} & $0.20 \pm 0.15$ & Nuclear dependence of 150 GeV $\pi - A$ D-Y cross sections \\
\end{tabular}
\end{center}
\end{table}

\newpage
\begin{figure}
  \begin{center}
    \mbox{\epsfysize=7.5in\epsffile{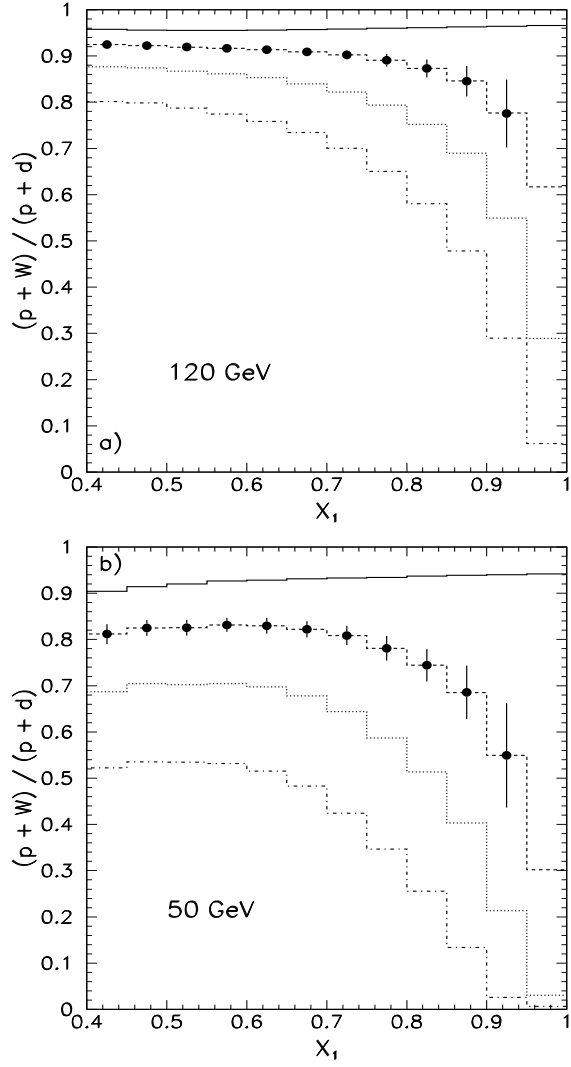}}
  \end{center}
\caption{ Calculated $(p+W) / (p+d)$ D-Y ratios at 50 GeV and 120 GeV
proton beam energies. 
The solid, dashed, dotted, and dash-dotted curves correspond
to  a linear partonic energy loss rate of 0.0, 0.1, 0.25, 0.5
GeV/fm, respectively. Also shown in the figure are the expected
statistical errors for $(p+W) / (p+d)$ ratios               
in a 60-day run for $p+W$ and $p+d$ each.}
\label{partonfig1}
\end{figure}

\begin{figure}
  \begin{center}
    \mbox{\epsfysize=7.5in\epsffile{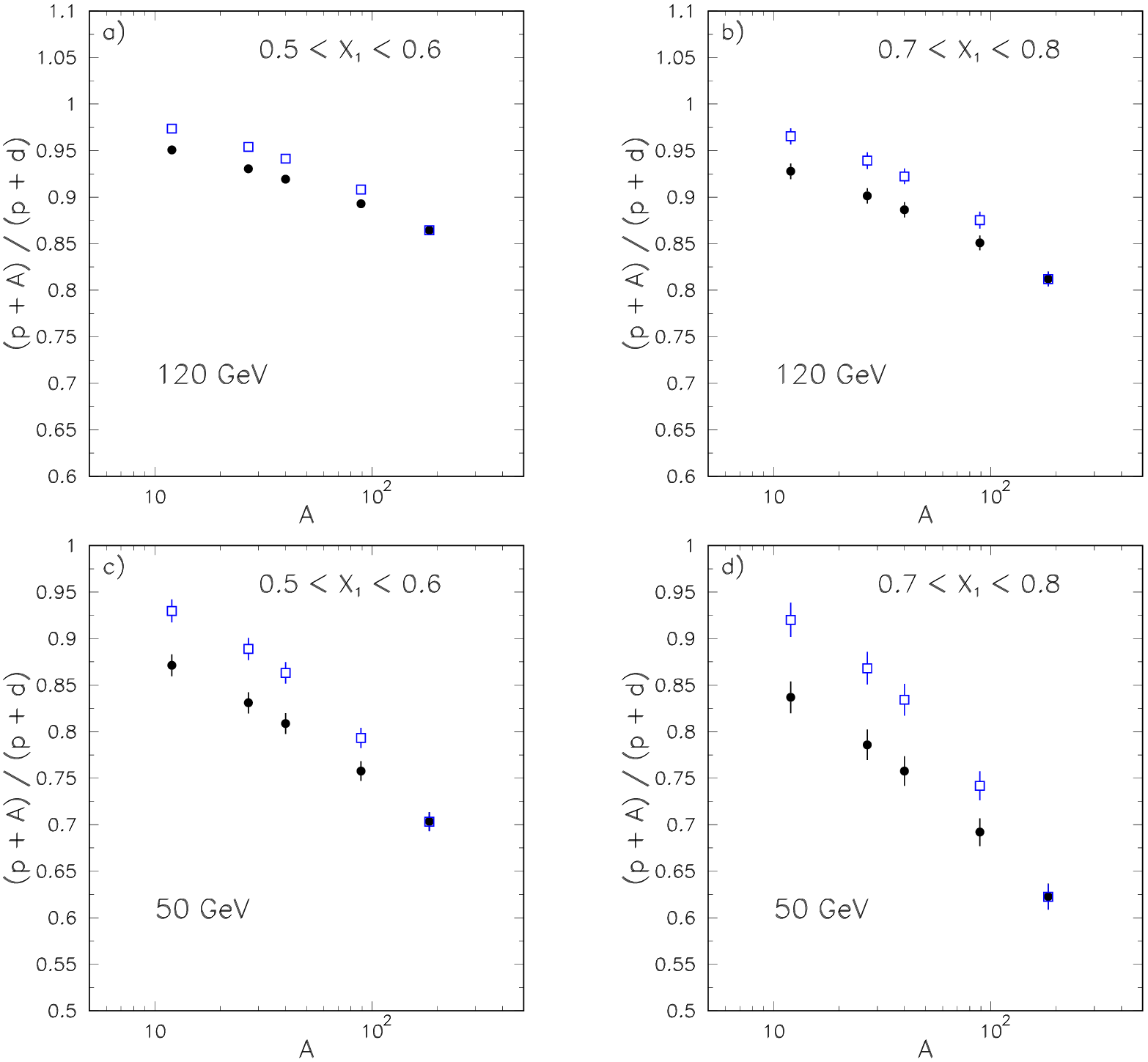}}
  \end{center}
\caption{Calculated D-Y $(p+A) / (p+d)$ ratios with different
assumption of the path-length dependence of partonic energy loss.
Solid circles correspond to 
a linear partonic energy loss rate of 0.25 GeV/fm.
The open squares correspond to a quadratic path-length dependence
of the partonic energy loss.
The statistical errors were calculated assuming a 60-day run for 
each target.}
\label{partonfig2}
\end{figure}


\begin{references}

\bibitem{gyulassy94} M. Gyulassy and X.N. Wang, Nucl. Phys. {\bf B420}, 
583 (1994).

\bibitem{zakharov96} B.G Zakharov, JETP Lett. {\bf 63}, 952 (1996);
{\bf 65}, 615 (1997).

\bibitem{baier97} R. Baier {\em et al.}, Nucl. Phys. {\bf B484}, 265 
(1997).

\bibitem{baier98} R. Baier {\em et al.}, Nucl. Phys. {\bf B531}, 403 
(1998).

\bibitem{baier01} R. Baier {\em et al.}, JHEP {\bf 0109}, 33 (2001).

\bibitem{yang98} Jian-Jun Yang and Guang-Lie Li, Eur. Phys. J. {\bf C5}, 
719 (1998).

\bibitem{guo00} Xiaofeng Guo and Xin-Nian Wang,  Phys. Rev. Lett. {\bf 85}, 
3591 (2000).

\bibitem{johnson02} M.B. Johnson {\em et al.}, Phys. Rev. {\bf C65}, 025203 (2002).

\bibitem{wang98} Xin-Nian Wang, Phys. Rev. Lett. {\bf 81}, 2655 (1998).

\bibitem{vasiliev99} M.A. Vasiliev {\em et al.}, Phys. Rev. Lett. {\bf 83}, 
2304 (1999).

\bibitem{johnson01} M.B. Johnson {\em et al.}, Phys. Rev. Lett. {\bf 86}, 
4483 (2001).

\bibitem{arleo02} Francois Arleo, Phys. Lett. {\bf B532}, 231 (2002).

\bibitem{wang02} Enke Wang and Xin-Nian Wang, hep-ph/0202105.

\bibitem{brodsky93} S. J. Brodsky and P. Hoyer, Phys. Lett. {\bf B298},
165 (1993).

\bibitem{alde90} D. M. Alde {\em et al.}, Phys. Rev. Lett. {\bf 64}, 2479 (1990).

\bibitem{pat99} P. L. McGaughey, J. M. Moss, J. C. Peng, Ann. Rev. Nucl. Part.
Sci. {\bf 49}, 217 (1999).

\bibitem{anderson79} K. J. Anderson {\em et al.}, Phys.
Rev. Lett. {\bf 42}, 944 (1979).

\bibitem{badier81} J. Badier {\em et al.}, Phys. Lett. {\bf B104}, 335 (1981); 
O. Callot {\em et al.}, Contribution to Moriond Workshop on Lepton Pair 
Production, Les Arcs (1981).

\bibitem{eskola98} K.J. Eskola, V.J. Kolhinen and P.V. Ruuskanen, Nucl. Phy. 
{\bf B535}, 351 (1998); K.J. Eskola, V.J. Kolhinen, and C.A. Salgado 
Eur. Phys. J. {\bf C9}, 61 (1999).

\bibitem{E906} D. Geesaman, P. Reimer et al., Fermilab proposal E906 (1999).

\bibitem{JHF} J. C. Peng et al., hep-ph/0007341 (2000); M. Asakawa et al.,
KEK-REPORT-2000-11.

\bibitem{mrst} A. D. Martin, R. G. Roberts, W. J. Stirling and R. S. Thorne,
Eur. Phys. J {\bf C4}, 463 (1998).

\end{references}
\end{document}